%% file: oja_template.tex
\documentclass[twocolumn]{openjournal}

\usepackage{xcolor}
\usepackage{textgreek}
\usepackage[utf8]{inputenc}
\usepackage[english]{babel}
\usepackage{longtable}
\usepackage{comment}

\usepackage{hyperref}
\hypersetup{
    unicode, 
    colorlinks=true,
    linkcolor=linkcolor,
    citecolor=linkcolor,
    filecolor=linkcolor,
    urlcolor=linkcolor,
}
\usepackage{color,colortbl}
\definecolor{linkcolor}{rgb}{0.0,0.3,0.5}
\usepackage{tensind}
\tensordelimiter{?}
\DeclareGraphicsExtensions{.bmp,.png,.jpg,.pdf}
\usepackage{verbatim}
\usepackage[normalem]{ulem}
\usepackage{orcidlink}
\usepackage{soul}

\usepackage{comment} 
\usepackage[detect-all]{siunitx} 
\usepackage{physics} 
\usepackage{acro} 
\acsetup{patch/longtable=false}
\usepackage{upgreek} 
\usepackage{subfigure}
\usepackage{booktabs} 
\usepackage{array}

\AtBeginDocument{\RenewCommandCopy\qty\SI}  
\DeclareSIUnit\ly{ly}
\DeclareSIUnit\Msun{M_{\odot}}
\DeclareSIUnit\Rs{R_{\mathrm{s}}}
\DeclareSIUnit\kms{km\,s^{-1}}

\input{glossary.tex}

\graphicspath{ {./figs/} }

\begin{document}

\title{Say Hello, Wave Goodbye: \\Gravitational Waves from Hyperbolic PBH-SMBH Interactions}

\author{Laura Burn\altaffilmark{1}\altaffilmark{*}\orcidlink{0009-0000-2396-5778}}
\author{Nelson Christensen\altaffilmark{2}\altaffilmark{\dag}\orcidlink{0000-0002-6870-4202}}
\author{Richard Easther\altaffilmark{1}\altaffilmark{\ddag}\orcidlink{0000-0002-7233-665X}}

\affiliation{\altaffilmark{1} Department of Physics, University of Auckland, Private Bag 92019, Auckland, New Zealand}
\affiliation{\altaffilmark{2} Université Côte d’Azur, Observatoire de la Côte d’Azur, CNRS, Laboratoire Artemis, 06300 Nice, France}

\altaffiltext{*}{\texttt{laura.burn@auckland.ac.nz}}
\altaffiltext{\dag}{\texttt{nelson.christensen@oca.eu}}
\altaffiltext{\ddag}{\texttt{r.easther@auckland.ac.nz}}

\begin{abstract}
Primordial black holes (PBHs) formed in the early Universe remain a viable dark matter candidate. Since dark matter is expected to be concentrated toward galactic centres, a significant population of PBHs would reside near a supermassive black hole (SMBH), providing a promising environment for gravitational wave emission. We consider PBHs with masses from ${10^{-15}}$ to ${10}\, {\mathrm{M}_\odot}$ on hyperbolic trajectories past an SMBH, that produce bursts of gravitational radiation during periastron passage. Sagittarius A* is an obvious location for these events, but we show that both M31* and M87* would generate similar signals, albeit at lower frequencies. We assess the detectability of both individual bursts and ``popcorn'' backgrounds, relative to LISA and the proposed $\upmu$Ares mission. Comparing these results to plausible central halo densities, we find that unbound PBHs are unlikely to yield detectable signals.  \vspace{12pt}
\end{abstract}

\vspace{1cm}

\maketitle

\section{Introduction}
\label{sec:intro}
\setcounter{footnote}{0}

The nature of dark matter is a key question in physics \citep{feng_dark_2010,undagoitia_dark_2016,bertone_history_2018,bertone_new_2018}. By the late 1990s, \ac{cdm} was established as the leading paradigm to describe the evolution of structure at all scales
\citep{peebles_large-scale_1982, blumenthal_formation_1984} and, according to the standard cosmological model, dark matter accounts for $\sim 27$\% of the mass-energy in the Universe \citep{collaboration_planck_2020}. However, direct detection experiments (e.g., \citealt{zhang_search_2022,lux-zeplin_collaboration_first_2023,xenon_collaboration_first_2023}) have not delivered evidence for popular particle models, such as axions \citep{peccei_cp_1977,weinberg_new_1978,wilczek_problem_1978} or \acp{wimp} (e.g., \citealt{arcadi_waning_2018,roszkowski_wimp_2018}), increasing interest in alternative candidates.

In this context, \acp{pbh} are receiving renewed attention as a dark matter candidate \citep{zeldovich_hypothesis_1967,hawking_gravitationally_1971,carr_primordial_2026}. In addition, \acp{pbh} might account for the surprisingly luminous early galaxies observed by the James Webb Space Telescope \citep{hutsi_did_2022}. Moreover, some of the black holes detected by the \ac{lvk} collaboration \citep{collaboration_gwtc-40_2025,LIGOScientific:2026wfs} are possibly consistent with a primordial origin \citep{bird_did_2016, ligo_scientific_collaboration_ligovirgokagra_2025}.

\Acp{pbh} are assumed to form in the early Universe either when primordial overdensities undergo gravitational collapse as they (re)enter the horizon \citep{harada_threshold_2013} or from overdensities initially generated by nonlinear sub-horizon processes, such as string networks, phase transitions, or parametric resonance \citep{carr_primordial_2022}. The mass range $\qtyrange{e-15}{e-12}{\Msun}$ is currently unconstrained, but above it there are non-trivial limits on any \ac{pbh} population. These constraints are derived from a variety of observations, including stellar microlensing \citep{niikura_constraints_2019} and the \ac{lvk} results \citep{andres-carcasona_constraints_2024}; see \citet{carr_primordial_2026} for a detailed review. Conversely, if \acp{pbh} emit Hawking radiation at the expected rate, those with masses below approximately $\qty{e-15}{\Msun}$ would have evaporated.\footnote{It has been argued that small \acp{pbh} are stabilised by the `memory-burden' induced by the information they carry, opening up the mass window below $\qty{e-15}{\Msun}$ \citep{dvali_memory_2024}.}

There is strong evidence that a \ac{smbh} resides at the centre of most large galaxies \citep{kormendy_coevolution_2013}. Since the density of dark matter is greatest at the centres of galaxies \citep{salucci_distribution_2019}, if \acp{pbh} constitute some or all of the dark matter, it follows that \ac{pbh}-\ac{smbh} interactions would be relatively common. Consequently, gravitational waves produced via these interactions could provide a distinctive signature of \ac{pbh} dark matter.

Thanks to its proximity, \ac{pbh} interactions with the Milky Way's central black hole, \ac{sgr} \citep{collaboration_polarimetry_2023}, have been the focus of most analyses of this possible signal. Previous work includes \citet{wang_searching_2020}, \citet{kuhnel_waves_2020}, and \citet{bondani_gravitational_2024}, which consider \acp{pbh} in bound (i.e., circular and elliptical) orbits around \ac{sgr} or \ac{pbh} binaries in its vicinity \citep{garcia-bellido_gravitational_2017, afroz_gravitational_2025}. However, since most \acp{pbh} in galactic centres would not be bound to the \ac{smbh}, we would expect multiple hyperbolic encounters.

The energy lost in a single hyperbolic encounter between an \ac{smbh} with mass $ M_{\mathrm{S}}$ and a \ac{pbh} with mass $M_{\mathrm{P}}$ is given by \citet{turner_gravitational_1977}:
\begin{equation}
    \delta E = \frac{8}{15} \frac{ G^{7/2}}{c^5} \frac{(M_{\mathrm{S}} + M_{\mathrm{P}})^{1/2} (M_{\mathrm{S}} M_{\mathrm{P}})^{2}}{r_{\mathrm{min}}^{7/2}} g(e) \, , \label{eq:energy}
\end{equation}
where $g(e)$ depends only on the eccentricity $e$. It is convenient to specify the distance of closest approach $r_{\mathrm{min}}$ in units of the Schwarzschild radius of the \ac{smbh} $R_{s}$, such that $r_{\mathrm{min}} = \hat  r_{\mathrm{min}} (2 G M_{\mathrm{S}}   /c^2 )$, and making the obvious approximation in the numerator gives
\begin{equation}
    \delta E = \frac{1}{15\sqrt{2}} \frac{c^2}{\hat r_{\mathrm{min}}^{7/2}} \frac{M_{\mathrm{P}}^2}{ M_{\mathrm{S}}} g(e) \, . \label{eq:energy2}
\end{equation}
Since  $\delta E$ scales as $M_{\mathrm{P}}^2$, low-mass \acp{pbh} are less likely to be captured into orbits around an \ac{smbh} by gravitational wave emission, and are thus more likely to undergo hyperbolic flybys. These would produce bursts of gravitational radiation during periastron passage \citep{berryExtrememassratioburstsExtragalacticSources2013}. In significant numbers, these passes would produce a popcorn background; an incoherent superposition of signals too weak to detect individually \citep{coward_detection_2006,Meacher:2015iua,romano_detection_2017,Christensen:2018iqi}.

\ac{smbh} event horizons  in large galaxies range from light minutes through to light hours in size. A \ac{pbh} making a close periastron passage moves at a significant fraction of the speed of light, so these lengths implicitly set the frequency range of the resulting gravitational waves, i.e., from a few $\mu$Hz to $\sim\qty{0.1}{Hz}$, which is primarily the domain of  space-based interferometers. In  particular, the \ac{lisa}  will operate in the $\qtyrange{0.1}{100}{mHz}$ frequency band \citep{amaro-seoane_laser_2017}, and analyses of \acp{pbh} of various masses interacting with \ac{sgr} typically focus on detection with \ac{lisa}. 

Furthermore, the proposed $\upmu$Ares detector would be sensitive to microhertz frequencies \citep{sesana_unveiling_2021}, which opens the door to \ac{pbh} interactions with M87*, the \ac{smbh} of the elliptical galaxy M87. M87* is $\sim 10^3$ times more massive than \ac{sgr} \citep{collaboration_first_2019}, as well as $\sim 10^3$ times more distant. We also consider the Andromeda galaxy (M31), the nearest large spiral galaxy to the Milky Way, whose central \ac{smbh} (M31*) is intermediate in both mass and distance between \ac{sgr} and M87*. Since strain only decreases linearly with distance, and because we expect more \ac{pbh}-\ac{smbh} interactions with a larger black hole, we consider all three galaxies as possible interaction sites in what follows.

Before launching into quantitative calculations, the form of $\delta E$ can be used to extract qualitative expectations. The energy injected into gravitational waves per encounter scales as $M_{\mathrm{P}}^2$. On the other hand, for a given density and velocity distribution of \acp{pbh}, the number of encounters, and thus the encounter rate $\Gamma$, grows linearly with  $1/M_{\mathrm{P}}$. We will see that the \ac{snr} scales with $\sqrt{\Gamma}$, so this signal will be more detectable at the top of the \ac{pbh} mass range. 

Increasing the \ac{smbh} mass with $M_{\mathrm{P}}$ fixed sees $\delta E$ decrease linearly with $M_{\mathrm{S}}$, but the energy flux  scales as $1/M_{\mathrm{S}}^2$, given that the transit-time grows with the physical size of the system. The characteristic frequency is inversely proportional to the transit time, and the energy carried by a gravitational wave scales with the square of its frequency. As a result, the expected strain in the source frame is independent of $M_{\mathrm{S}}$ for a single encounter. Individual signals from M31* and M87* will be attenuated by their distance, but event rates will be higher than for \ac{sgr}, given their greater cross-sectional area for a given density of \acp{pbh}.  

In what follows, we compute the density of \acp{pbh} required to give rise to a popcorn gravitational wave background within the sensitivity range of future space-based detectors arising from hyperbolic passes by \acp{pbh}. We also investigate the probability of detecting a burst-like signal from a single hyperbolic encounter. We focus on \ac{sgr}, M31* and M87*, but our results generalise to other possible scenarios. Obviously, any gravitational wave signal from \ac{pbh}-\ac{smbh} interactions will depend on both the \ac{pbh} contribution to dark matter and the dark matter distribution at the centres of galaxies. Beyond the hypothetical nature of \ac{pbh} dark matter itself, the dark matter density at halo centres is poorly constrained 
\citep{burkert_structure_1995, salucci_dark_2000}. Consequently, our goal is to make an estimate rather than an overly detailed calculation.

The structure of this paper is as follows. Section~\ref{sec:darkmatterhalo} delineates the halo profiles for our three target systems; Section~\ref{sec:geometry_event_rate} summarises the orbital geometries, and Section~\ref{sec:spectrum} computes the spectra. The detectability of these signals is assessed in Section~\ref{sec:detect} and we conclude in Section~\ref{sec:conclusion}.
\begin{table*}[t]
\begingroup 
    \setlength{\tabcolsep}{5pt} 
    \renewcommand{\arraystretch}{1.5} 
    \setlength{\extrarowheight}{2pt}
        \centering
    \begin{tabular*}{\textwidth}{@{\extracolsep{\fill}}lccccccl}
        & \multicolumn{2}{c}{Milky Way} & \multicolumn{2}{c}{M31} & \multicolumn{2}{c}{M87} & \\
        \hline \hline
        Profile 
        & $r_s\, (\mathrm{kpc})$ & $\rho_s\, (M_{\odot}\,\mathrm{kpc}^{-3})$ 
         & $r_s\, (\mathrm{kpc})$ &$\rho_s\, (M_{\odot}\,\mathrm{kpc}^{-3})$
         & $r_s\, (\mathrm{kpc})$ &$\rho_s\, (M_{\odot}\,\mathrm{kpc}^{-3})$
        & Reference \\
        \hline \hline
        NFW 
        & $8.1$ & $5.2 \times 10^7$
        & $16.5$ & $1.1 \times 10^7$
        & 51.0 & $1.1 \times 10^7$
        & \begin{tabular}{l} \cite{lin_dark_2019}; \cite{tammStellarMassMap2012}; \\ \cite{li_discrete_2020} \\ \end{tabular} \\

        $M_{\mathrm{S}}-\sigma$ 
        & 24.7 & $3.7 \times 10^6$
        & 73.3 & $2.0 \times 10^6$
        & 242.8 & $1.1 \times 10^6$
        & \begin{tabular}{l} \cite{croton_simple_2009}; \cite{dutton_cold_2014} \\ \end{tabular} \\  

        Burkert 
        & 7.8 & $5.2 \times 10^7$
        & 9.1 & $3.7 \times 10^7$
        & 91.7 & $6.9 \times 10^6$
        & \begin{tabular}{l} \cite{lin_dark_2019}; \cite{tammStellarMassMap2012}; \\ \cite{jusufi_black_2019} \\ \end{tabular} \\
        \hline \hline 
        \end{tabular*}  
        \caption{Characteristic parameters for the dark matter halo density profiles considered here.}
        \label{tb:profiles}
\endgroup
\end{table*}
\section{Dark Matter Halo Density Profile}\label{sec:darkmatterhalo}
We define the \ac{pbh} population by positing a halo profile and taking the resulting density and velocity distribution near the centre. The halo density generically scales as $1/r$ at a large radial distance $r$, but the central regions are less understood. We can consider both the canonical \Ac{nfw} profile and the Burkert profile. The former has the general form 
\begin{equation}
    \rho_{\mathrm{NFW}}(r) = \rho_{s} \, \frac{r_s} {r \left( 1+\frac{r} {r_{s}} \right)^{2}}\,,
    \label{eq:nfw}
\end{equation}
where $r_{s}$ and $\rho_{s}$ are the scale radius and density, respectively \citep{navarro_structure_1996}. The Burkert profile is given by 
\begin{equation}
    \rho_{\mathrm{bur}}(r)=\frac{\rho_{s}r_{s}^{3}}{(r+r_{s})(r^{2}+r_{s}^{2})}\,,
\end{equation}
where $r_s$ and $\rho_s$ are now the core radius and central density, respectively. The Burkert profile was first proposed to explain the observed rotation curves of dwarf spiral galaxies \citep{burkert_structure_1995, salucci_dark_2000} and, unlike the \ac{nfw} profile, does not diverge at small $r$.  

\citet{lin_dark_2019} and \citet{tammStellarMassMap2012} give parameter values for the \ac{nfw} and Burkert profiles for the Milky Way and M31, respectively, while \citet{li_discrete_2020} and \citet{jusufi_black_2019} provided values for the M87 profiles. \citet{gondolo_dark_1999} suggest that the adiabatic growth of the \ac{smbh} can enhance the dark matter density in its immediate vicinity, resulting in a spiked distribution. However, any local feature is  bound to the \ac{smbh} and will not contribute to the hyperbolic encounters considered here. Conversely, if \acp{pbh} make a fractional contribution to the dark matter, we assume that their distribution matches the overall halo profile. 

The scale radius and density can also be estimated from the mass of the \ac{smbh}. The \ac{smbh} masses themselves are well-measured; for the Milky Way, $M_{\mathrm{S}}=\qty{4.3e6}{\Msun}$  \citep{collaboration_polarimetry_2023}, for M31, $M_{\mathrm{S}}=\qty{1.4e8}{\Msun}$ \citep{benderHSTSTISSpectroscopy2005} and for M87, $M_{\mathrm{S}}=\qty{6.5e9}{\Msun}$  \citep{collaboration_first_2019}. The empirical $M_{\mathrm{S}}-\sigma$ relation 
connects the \ac{smbh} mass to the velocity dispersion in the bulge $\sigma$, and is associated with the co-evolution of the \ac{smbh} and its host galaxy \citep{kormendy_coevolution_2013}. \citet{croton_simple_2009} connects the \ac{smbh} mass to the virial mass of the dark matter halo $M_{\mathrm{vir}}$ via the quasar luminosity function. Since we are making observations in the present epoch we can use this result in the $z=0$ limit, 
\begin{equation}
\begin{split}
&\log\left(\frac{M_{\mathrm{S}}}{10^{8}h^{-1}M_\odot}\right) = -2.66 +\\
& \qquad \qquad 1.39 \times \log\left[\gamma^{3} H_0\left(\frac{M_{\mathrm{vir}}}{10^{13}h^{-1}M_\odot}\right)\right]\,.
\end{split}
\end{equation}
As usual, the Hubble constant $H_0$ is written as $100\,h\,\mathrm{km/s\,Mpc^{-1}}$ and we set $h=0.674$ for definiteness \citep{planck_collaboration_planck_2020}. The ratio between the dark matter halo's circular velocity and the virial velocity is given by $\gamma$, which is of order unity \citep{porciani_cosmic_2004}. The $M_{\mathrm{S}}-\sigma$ relation and the concentration-mass relation was obtained by \citet{dutton_cold_2014} via fits to N-body simulations with Planck 2013 parameters \citep{collaboration_planck_2014}. This gives $r_s$ and $\rho_s$ for an \ac{nfw} halo as a function of $M_{\mathrm{S}}$. 

\begin{figure}[!t]
    \centering
    \vskip 8pt
    \includegraphics[width=\columnwidth]{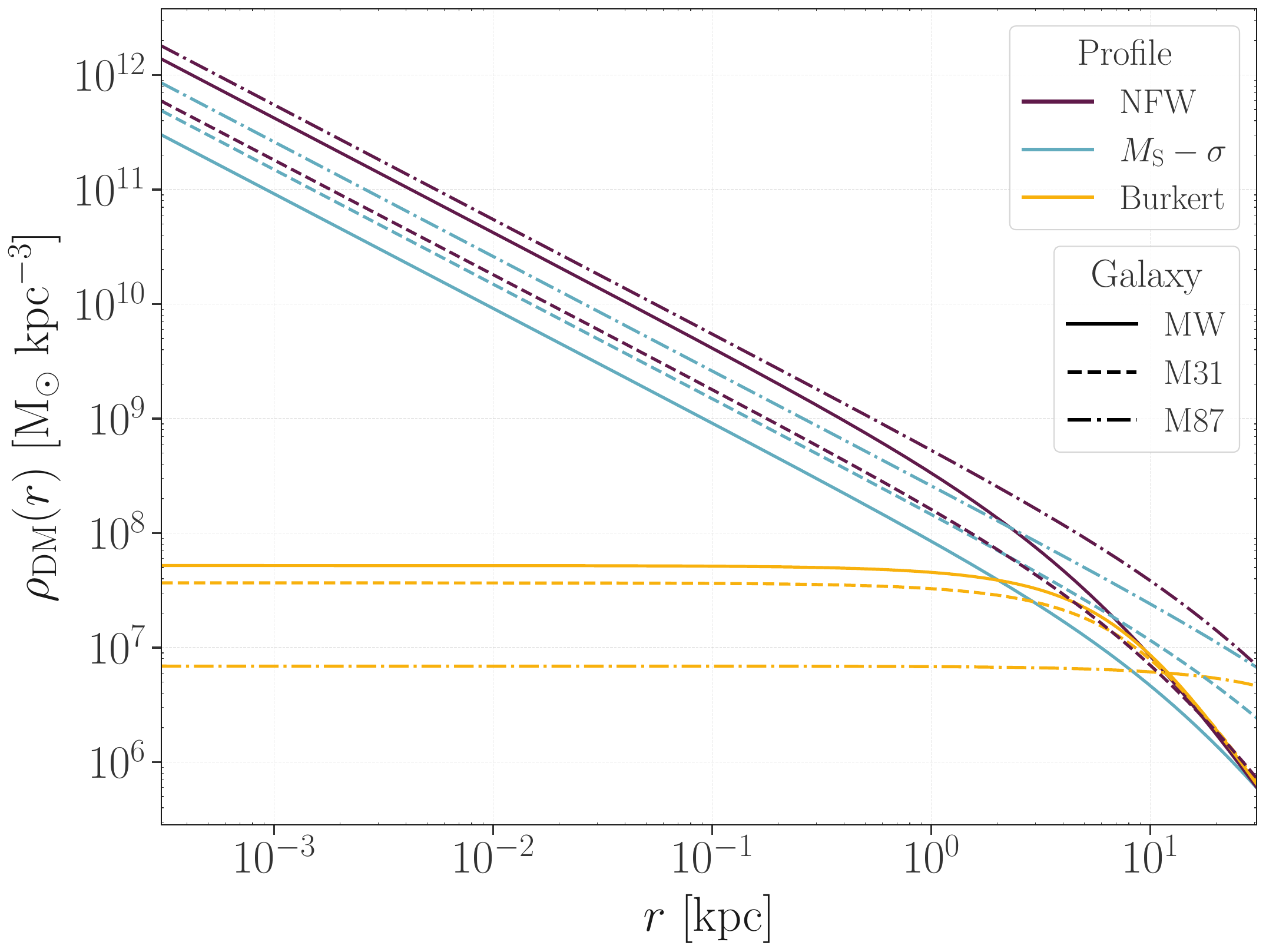}
    \caption{The dark matter halo density profiles as a function of radius for the Milky Way (solid), M31 (dashed) and M87 (dashed-dotted); purple represents the fitted \ac{nfw} profile, blue is an \ac{nfw} profile derived from the $M_{\mathrm{S}}-\sigma$ relation, and yellow is the Burkert profile.}
    \vskip 1pt
    \label{fig:darkmatterprofile}
\end{figure}

Parameters for the Milky Way, M31 and M87 halos from the $M_\mathrm{S}-\sigma$ relation are listed alongside direct estimates of the \ac{nfw} and Burkert profiles in Table~\ref{tb:profiles} and the profiles are plotted in  Fig.~\ref{fig:darkmatterprofile}. In what follows, we take the \citet{lin_dark_2019}, \citet{tammStellarMassMap2012} and \citet{li_discrete_2020} profiles as our reference points, so our results are effectively upper bounds on the possible gravitational wave signal from hyperbolic \ac{pbh}-\ac{smbh} interactions.  We note that the distribution of dark matter at the centres of galaxies is poorly constrained, and that there is also no guarantee that all halos have the same form of $\rho_{\mathrm{DM}}(r)$ in their central regions, which may be a function of their morphology and merger history. 

\section{Geometry and Event Rate}
\label{sec:geometry_event_rate}
The hyperbolic path of a \ac{pbh} passing a central \ac{smbh} is a standard result. %
\begin{figure}[!t]
    \centering
    \vskip 6pt
    \includegraphics[width=\columnwidth]{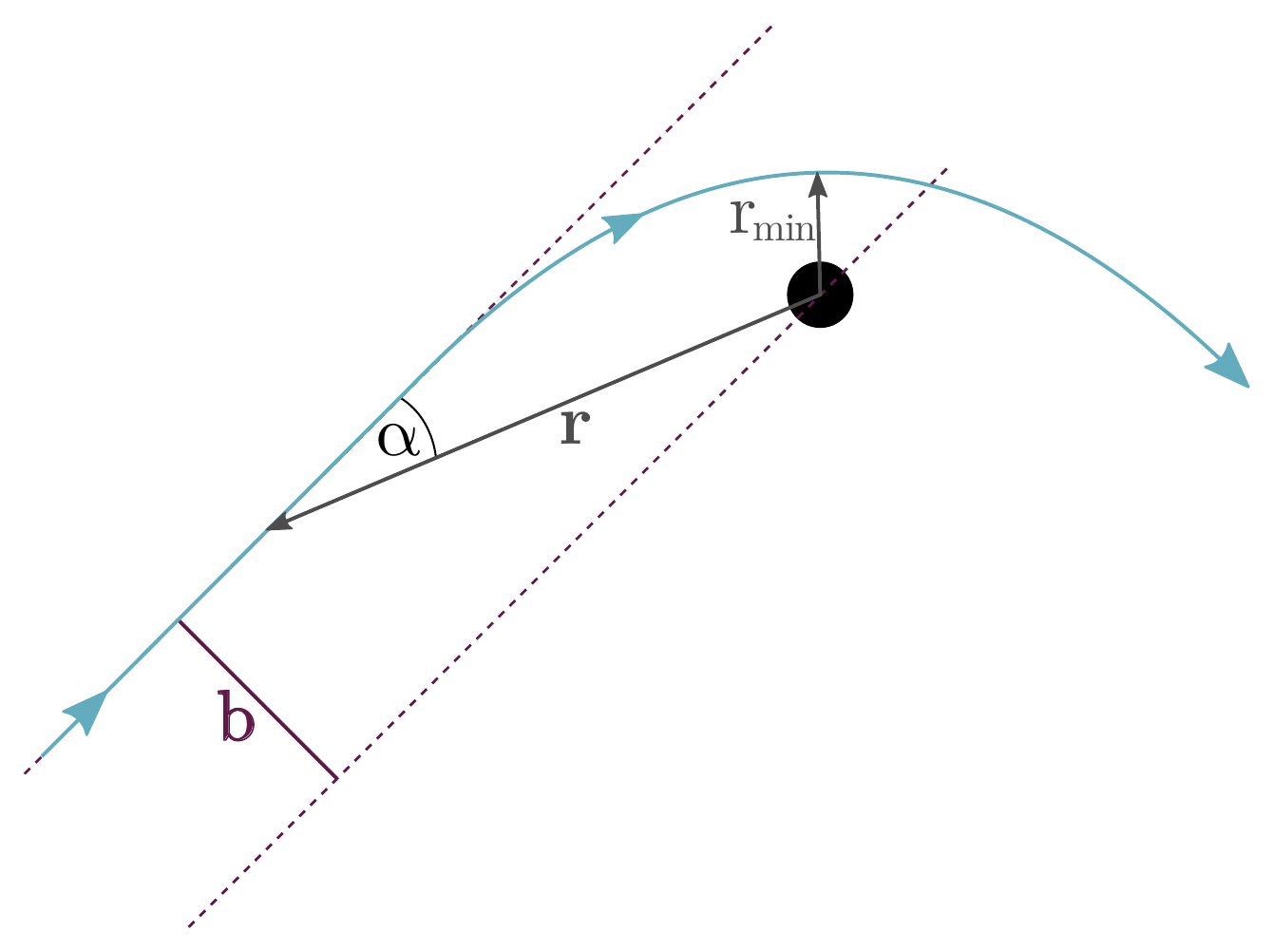}
    \caption{A hyperbolic trajectory (solid blue) of a \ac{pbh} passing an \ac{smbh} (black, filled circle), described by position $\textbf{r}$ and angle $\alpha$. The closest approach on this trajectory occurs at periastron $r_{\mathrm{min}}$. The impact parameter $b$ (solid purple) would have been the closest approach on the unmodified trajectory, without gravitational focusing.}
    \label{fig:hyperbolicgeometry}
\end{figure}
It is characterised by the impact parameter $b$ and the angle $\alpha$, which are related by
\begin{equation}
    b = r \sin \alpha\,,
\end{equation}
and is sketched in Fig.~\ref{fig:hyperbolicgeometry}.
As reviewed in Appendix \ref{ap:r_min}, the periastron distance is  
\begin{equation}
    r_{\mathrm{min}}= \sqrt{b^2 + \left(\frac{GM_{\mathrm{S}}}{v_0^2}\right)^2} - \frac{GM_{\mathrm{S}}}{v_0^2}\,,
    \label{eq:r_min_main}
\end{equation}
where $v_0$ is the asymptotic velocity. In the limit that $b$ is very large $b\approx r_{\mathrm{min}}$ and its motion is essentially linear. In the opposite limit, 
\begin{equation} 
\hat r_{\mathrm{min}} \approx \hat b^2\frac{v_0^2}{c^2} \, , 
\end{equation}
where $\hat b$ expresses $b$ in units of Schwarzschild radii. The latter result reflects the strong, velocity-dependent focusing of trajectories by the \ac{smbh}.

We frame the problem by imagining the \ac{smbh} in a uniform population of \acp{pbh} with a Maxwell-Boltzmann velocity distribution 
\begin{equation}
    f ( v ) \, \mathrm{d} v \equiv4\pi v^2\left( \frac{3} {2 \pi\, v_{\mathrm{rms}}^{2}} \right)^{3 / 2} \operatorname{e x p} \left(-\frac{3 \, v^{2}} {2 \, v_{\mathrm{rms}}^{2}} \right) \mathrm{d}v\,,
\end{equation}
where $v_{\mathrm{rms}}$ is the root mean square velocity. There is considerable uncertainty in the velocity distribution of dark matter halos \citep{helmi_streams_2020}; in what follows we set $v_{\mathrm{rms}} = \qty{100}{\kms}$, unless stated otherwise, and show in Sec.~\ref{sec:spectrum} that the specific value of $v_{\mathrm{rms}}$ has little impact on our conclusions.

We set the \ac{pbh} density to the halo density at a reference distance $r_i$, an order of magnitude larger than the characteristic gravitational focusing scale $c^2/v_0^2$, or equivalently, where $v_0$ is substantially larger than the \ac{smbh} escape velocity. We take $r_i=\qty{e2}{ly}$, $\qty{e3}{ly}$ and $\qty{e5}{ly}$ for the Milky Way, M31 and M87, respectively. Our overall strategy is to build a sample of \acp{pbh} making a close pass to the \ac{smbh} and then sum over these (and their associated rate) to produce the gravitational wave signals. 

We specify trajectories in terms of $v_0$ and $b$, with the former being directly drawn from the Maxwell-Boltzmann distribution. Since the gravitational wave signal scales as $r_{\mathrm{min}}^{-7/2}$, the vast majority of hyperbolic passes produce few gravitational waves. Consequently, we importance sample by restricting attention to trajectories for which $r_{\mathrm{low}}\leq r_{\mathrm{min}} \leq r_{\mathrm{high}}$. We take $r_{\mathrm{low}} = \qty{1}{\Rs}$, eliminating \acp{pbh} plunging directly into the \ac{smbh}, and set $r_{\mathrm{high}} = \qty{20}{\Rs}$, as we will see in Sec.~\ref{sec:spectrum} that the resulting flux at this radius is very small. Trajectories with impact parameters between $b$ and $b+db$ occupy an annular area $dA = 2\pi b \,db$, and we draw $b \in [b_{\mathrm{low}}, b_{\mathrm{high}}]$ from the normalised distribution
\begin{equation}
    f(b) = \frac{2b}{b_{\mathrm{high}}^2(v_0)-b_{\mathrm{low}}^2(v_0)}\, ,
\end{equation}
via Eq.~\ref{eq:r_min_main}. Trajectories are then weighted by their corresponding velocity-dependent cross-sectional area
\begin{equation}
    \sigma(v_0) = \pi[b_{\mathrm{high}}^2(v_0) - b_{\mathrm{low}}^2(v_0)]\,.
\end{equation}
The weights are thus
\begin{equation}
    w_i = \frac{n_{\mathrm{P}} }N v_0^{(i)} \sigma(v_0^{(i)})\,,
\end{equation}
where $N$ is the number of samples taken, $n_{\mathrm{P}}$ is the \ac{pbh} number density
\begin{equation}
    n_{\mathrm{P}} = \frac{f_{\mathrm{P}}\rho_{\mathrm{DM}}}{M_{\mathrm{P}}}\,,
\end{equation}
and $f_{\mathrm{P}}$ is the \ac{pbh} dark matter fraction, assuming a monochromatic mass function.

The resulting total encounter rate of hyperbolic \acp{pbh} with an \ac{smbh} for the given periastron range is the sum 
\begin{equation}
\Gamma = \sum_{i=1}^N w_i = \frac{n_{\mathrm{P}}}{N}\sum_{i=1}^N v_0^{(i)} \sigma(v_0^{(i)})\,.
\end{equation}
We worked with $N=10^7$ and tested that our results were independent of $N$. We can check the self-consistency of this result by noting that if we choose a very large sphere of radius $b_{\mathrm{max}}$, then the flux through it is $\pi n_{\mathrm{P}} \langle v \rangle  b_{\mathrm{max}}^2$, while $f(b)$ is reduced by a factor of $\pi b_{\mathrm{max}}^2$. The factor of 4 from the spherical surface area is accounted for by averaging over the angular directions, and the sum over $i$ recovers the average speed. 
\begin{figure}[!t]
    \centering
    \includegraphics[width=\columnwidth]{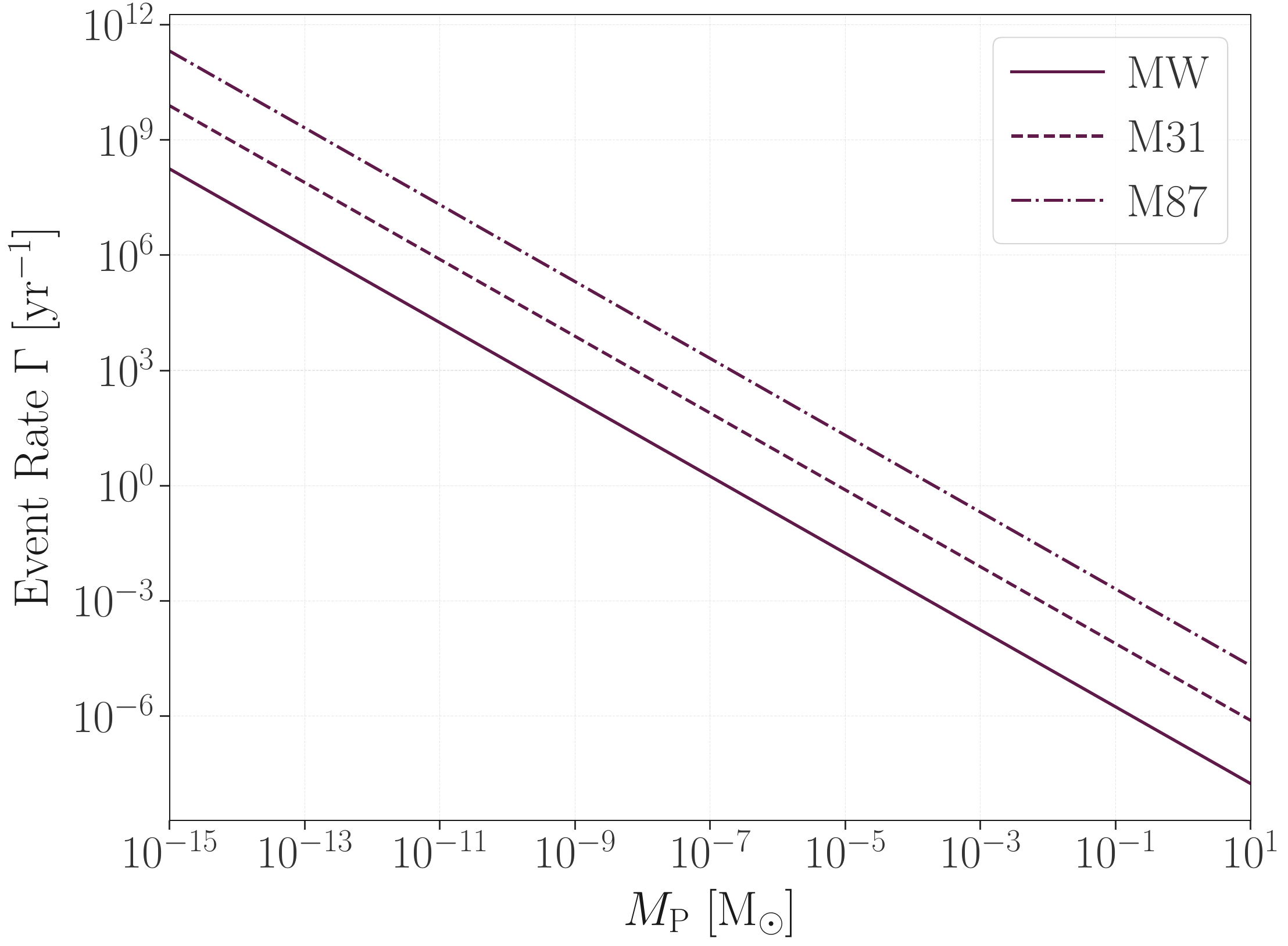}
    \caption{The number of hyperbolic \acp{pbh} passing within $\qtyrange{1}{20}{\Rs}$ of the host \ac{smbh} in a year for \acp{pbh} in the mass range $\qtyrange{e-15}{e1}{\Msun}$, assuming an \ac{nfw} profile with values from Table~\ref{tb:profiles}.}
    \label{fig:eventrate}
\end{figure}

Fig.~\ref{fig:eventrate} shows the nominal annual event rate for our three systems, assuming our canonical \ac{nfw} densities and $1\leq \hat r_{\mathrm{min}} \leq 20$, as a function of the \acp{pbh} mass. Since the fraction of asteroid mass \acp{pbh} is unconstrained, encounter rates exceeding one per second are possible for all three galaxies. However, for larger masses $f_{\mathrm{P}} < 1$ \citep{carr_constraints_2020}, and these numbers should be regarded as an upper bound. 

\begin{figure}[!t]
    \centering
    \includegraphics[width=\columnwidth]{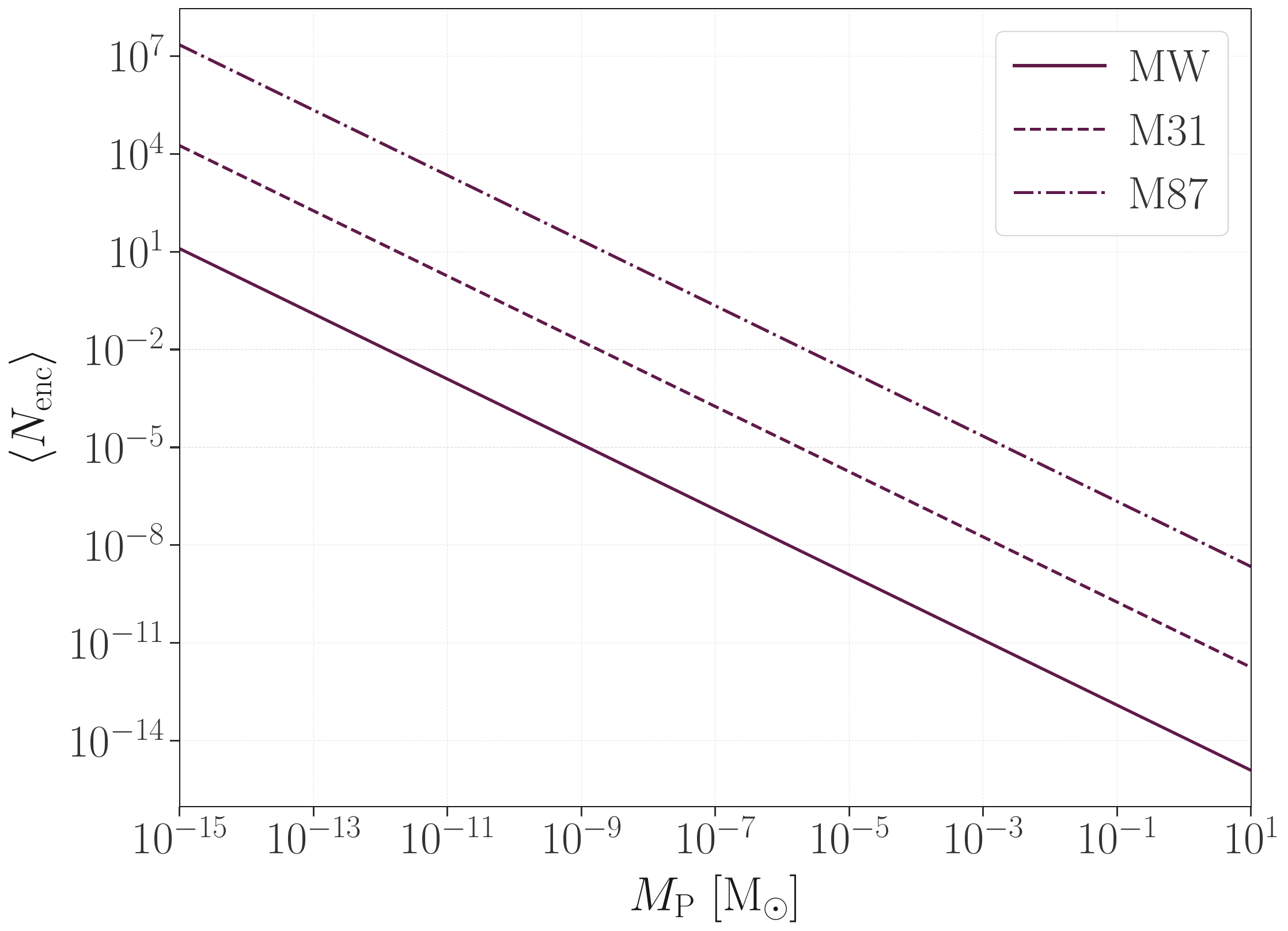}
    \caption{The average number of \ac{pbh} encounters per periastron bin during one Schwarzschild crossing time $R_s / c$ as a function of \ac{pbh} mass.}
    \label{fig:average_binned_rate}
\end{figure}

For small periastron distances, the encounter is in a strong gravitational-focusing regime and $b^2 \sim r_{\mathrm{min}}$, so $\sigma$ and thus $\Gamma$, scale linearly with $r_{\mathrm{min}}$. Consequently, the differential event rate $d\Gamma/dr_{\mathrm{min}}$ is approximately constant for small $r_{\mathrm{min}}$. Fig.~\ref{fig:average_binned_rate} shows the average number of encounters for our three systems as a function of the \ac{pbh} mass, again with $f_\mathrm{P}=1$, relative to the light-crossing time of the Schwarzschild radius for each \ac{smbh}. Each event lasts a few times longer than this, so this effectively measures the potential number of simultaneous events in each system. We see that M87* could experience approximately $10^6$ more events at any given moment than \ac{sgr}, but that all three systems could see multiple overlapping signals if the \ac{pbh} mass is at the bottom of its range.  

\begin{figure}[!t]
        \centering
    	\includegraphics[width=\columnwidth]{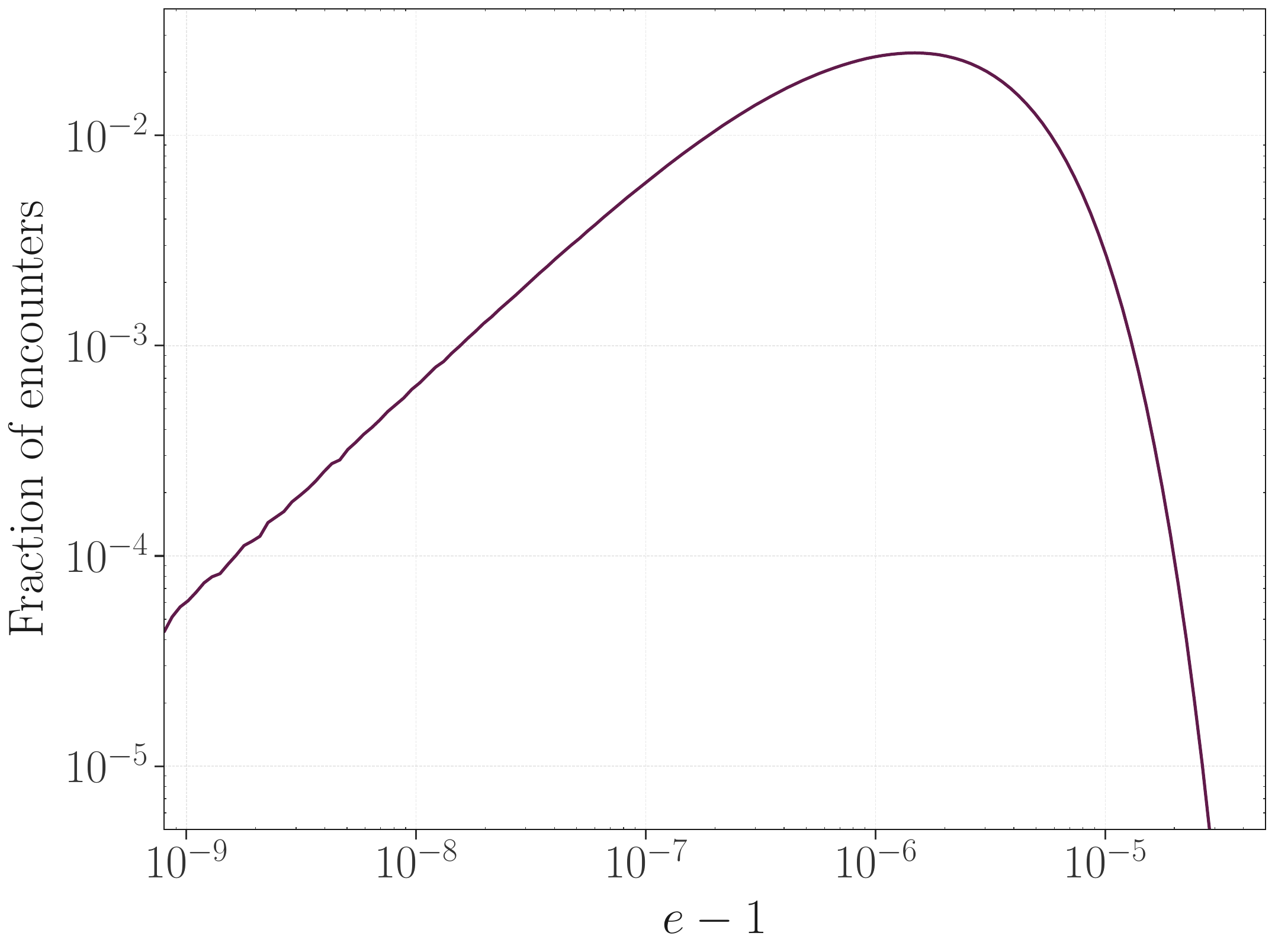}
    	\caption{The fraction of \ac{pbh} encounters as a function of $e-1$, where $e$ is the eccentricity. The distribution is normalised by the total number of encounters and averaged over the three galaxies. The encounters are predominantly parabolic, with $e \simeq 1$.}
        \vskip 6pt
        \label{fig:e_fraction}
\end{figure}

For hyperbolic orbits $e > 1$ and, as summarised in Appendix \ref{ap:r_min}
\begin{equation}
    e = \sqrt{\frac{(v_0^2 b)^2}{(GM_{\mathrm{S}})^2} + 1}\,.
    \label{eq:ecc}
\end{equation}
In the strong gravitational-focusing regime, $e$ is very close to unity and the encounters are close to parabolic. Fig.~\ref{fig:e_fraction} shows the distribution of $e-1$, normalised by the total number of encounters –- this is the same for each galaxy for a given $v_{\mathrm{rms}}$. We see that the vast majority of encounters are marginally unbound and close to parabolic.

\newpage
\section{Gravitational Wave Energy Spectrum}
\label{sec:spectrum}

The actual characteristic timescale of the encounter $\tau$ sets the peak frequency $f_{\mathrm{peak}}$ of the emitted gravitational wave signal, $f_{\mathrm{peak}}\sim 1/\tau$. While the Schwarzschild crossing time $R_s / c$ is a lower bound on $\tau$, as noted above, it is estimated accurately by \citet{turner_gravitational_1977},
\begin{equation}
\tau = \frac{r_{\mathrm{min}}^{3/2}}{\sqrt{G(M_{\mathrm{S}}+M_{\mathrm{P}})(e+1)}}\,.
\label{eq:tau}
\end{equation}
The encounter times for \acp{pbh} with \ac{sgr} with small periastron values are on the order of hours, placing $f_{\mathrm{peak}}$ near the frequency sensitivity of both \ac{lisa} and $\upmu$Ares. M31* is $\sim 10^2$ times more massive and $f_{\mathrm{peak}}$ is shifted to lower frequencies where $\upmu$Ares is more sensitive than \ac{lisa}. For M87*, which is $\sim 10^3$ times more massive than \ac{sgr}, encounters take days. For the corresponding frequencies, \ac{lisa} is dominated by residual acceleration noise in the test-masses \citep{babakLISASensitivitySNR2021}, greatly reducing its sensitivity. However, the longer arm lengths of $\upmu$Ares make it sensitive to frequencies below those accessible to \ac{lisa} and therefore will achieve a higher \ac{snr} for encounters around M87* than \ac{lisa}. The \ac{pbh}-\ac{smbh} interactions from these three systems thus span the majority of the frequency space covered by future space-based gravitational wave detectors.

The spectral energy density of a single hyperbolic encounter is expressed as combinations of modified Bessel functions of the second kind $K_{n}$ \citep{turner_gravitational_1977}
\begin{equation}
\begin{aligned}
\frac{\dd (\delta E)}{\dd f} &= 
\frac{16}{15}\frac{G^{7/2}}{c^{5}}
\frac{({M_{\mathrm{S}} + M_{\mathrm{P}}})^{1/2}(M_{\mathrm{S}}M_{\mathrm{P}})^2}{r_{\mathrm{min}}^{7/2}}
\,\tau\sqrt{e} \\
\quad &\times \Big\{
12 [u^{2}K_{2}(u) - uK_{1}(u)]^{2} \\
\qquad &+ 3 [2u^{2}K_{1}(u) + uK_{0}(u)]^{2}
+ u^{2}K_{0}^{2}(u)
\Big\}\,,
\end{aligned}
\end{equation}
where we define the dimensionless frequency parameter $u = \omega\tau = 2\pi f \tau$. The gravitational wave spectral luminosity is then 
\begin{equation}
    \frac{\dd L}{\dd f} = \frac{1}{\tau} \frac{\dd (\delta E)}{\dd f}\,.
\end{equation}
Assuming isotropy, the emitted energy is spread over the spherical surface of radius $d$, giving a detected spectral flux
\begin{equation}
\mathcal{F} ( f )=\frac{1} {4 \pi d^{2}} \frac{\dd L} {\dd f}\,. 
\end{equation}
For \ac{sgr}, $d = \qty{8.23}{kpc}$ \citep{collaboration_polarimetry_2023}, for M31*, $ d = \qty{765}{kpc}$ \citep{riessCepheidPeriodLuminosityRelations2012}, and for M87*, $d = \qty{16.4}{Mpc}$ \citep{bird_inner_2010}. The total spectral flux $\mathcal{F}_{\mathrm{total}}(f)$ from all encounters is estimated by taking the mean flux and scaling by the event rate $\Gamma$
\begin{equation}
\mathcal{F}_{\mathrm{total}}(f) = \frac{\Gamma}{N} \, \sum_{i=1}^{N}\mathcal{F}_i(f) \,.
\end{equation}
\begin{figure}[!t]
        \centering
    	\includegraphics[width=\columnwidth]{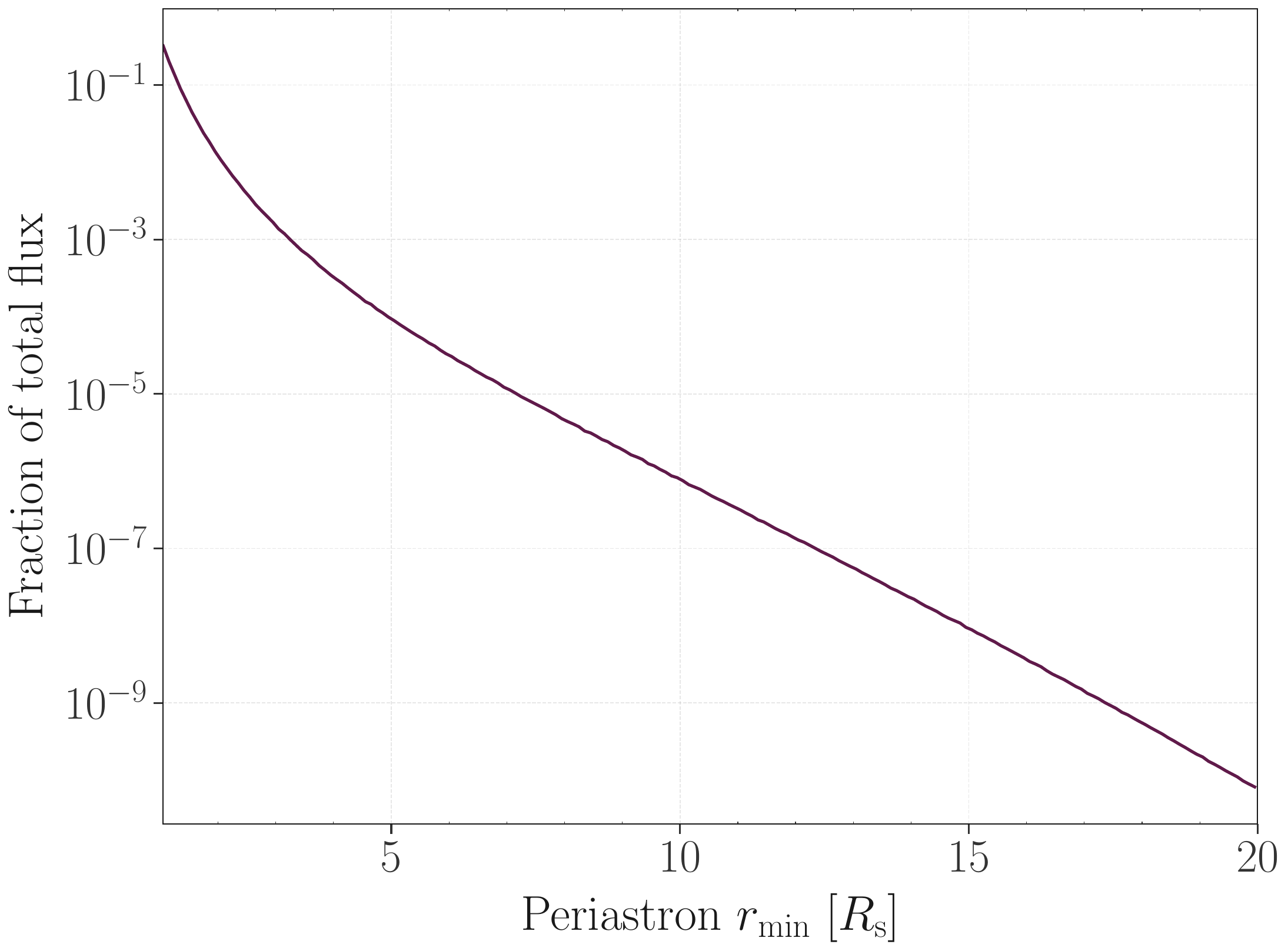}
    	\caption{The fraction of the total spectral flux contributed by \ac{pbh} encounters at different periastron distances. The flux is integrated over frequency between the minimum and maximum values of $f_{\mathrm{peak}}$ in the contributing population and normalised by the total flux. The distribution is averaged over the three galaxies. The closest \ac{pbh} encounters constitute most of the spectral flux.}
        \label{fig:flux_per_r}
\end{figure}
In a population of \ac{pbh} encounters, those with the smallest periastrons will dominate the total spectral flux. Fig.~\ref{fig:flux_per_r} shows the normalised spectral flux as a function of logarithmic periastron distance; since this is a relative intensity, the \ac{smbh} and \ac{pbh} mass falls out of the calculation. This plot also justifies our choice to limit attention to periastron distances between $\qty{1}{\Rs}$ and $\qty{20}{\Rs}$, as most of the total spectral flux in this case comes from encounters with $r_{\mathrm{min}} \leq 2 R_\mathrm{s}$. These events are rare and, as discussed in Sec.~\ref{sec:detect}, we should treat these as individual events, rather than as a popcorn background.

The signal strength is conventionally quantified via the characteristic strain, since the height of the source above the detector noise curve is directly related to the \ac{snr}. However, the characteristic strain is most applicable to long-lived sources such as inspirals, where signal accumulates over many cycles. We instead use the square root of the \ac{psd} $S_h(f)$, commonly referred to as the \ac{asd}, and is given by \citet{moore_gravitational-wave_2015},
\begin{equation}
    \sqrt{S_h(f)} = 2f^{1/2}|\tilde{h}(f)|\,,
\end{equation}
where $|\tilde{h}(f)|$ is the Fourier transform of the strain
\begin{equation}
|\tilde{h}(f)| = h(f) = \sqrt{\frac{4 G}{\pi c^3} \frac{\mathcal{F}_{\mathrm{total}}(f)}{f^2}}\,.
\end{equation}
The \ac{asd} characterises the frequency-dependent amplitude of the gravitational wave, giving a strain amplitude per $\sqrt{\mathrm{Hz}}$ and allows direct comparison with detector sensitivity curves. We use the event rates in Fig.~\ref{fig:eventrate} to evaluate the \ac{asd} of a popcorn gravitational wave background and assess detectability in  \ac{lisa} and $\upmu$Ares, assuming the canonical \ac{nfw} profile for each galaxy and $f_{\mathrm{P}}=1$. 
Since the duration of an encounter depends on $v_0$ and $r_{\mathrm{min}}$ (Eq.~\ref{eq:tau}), encounters will peak across a range of frequencies. We evaluate the \ac{asd} for the gravitational wave background over the frequency range set by the minimum and maximum encounter durations for each \ac{smbh} and \ac{pbh} mass. 
\begin{figure}[!t]
    \centering
    \includegraphics[width=\columnwidth]{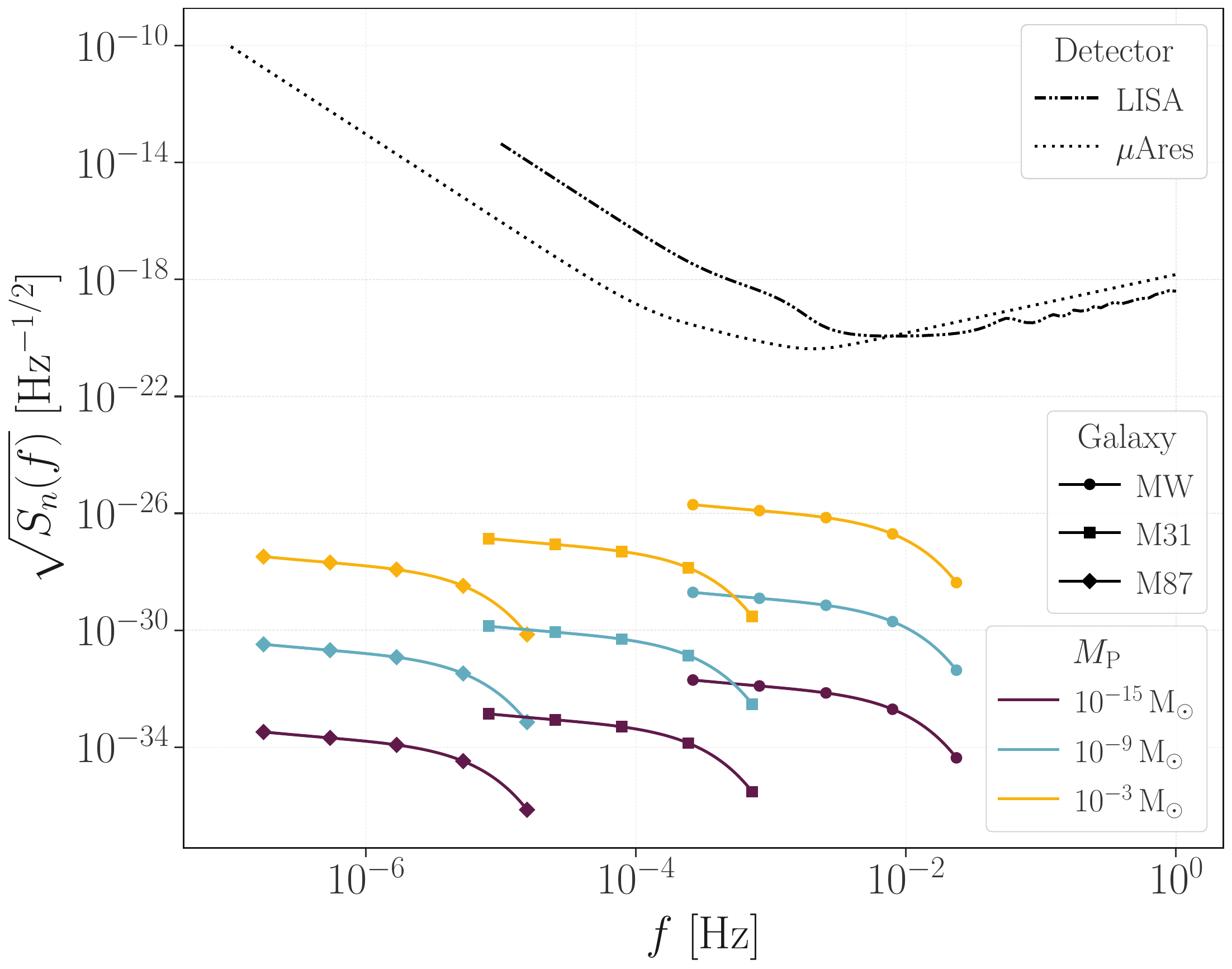}
    \caption{The frequency-dependent \ac{asd} for a popcorn gravitational wave background produced by a population of \acp{pbh} on hyperbolic orbits passing by \ac{sgr} (circles), M31* (squares), and M87* (diamonds) with a periastron distance between $\qtyrange{1}{20}{\Rs}$. The \ac{asd} for three \ac{pbh} masses are displayed; $\qty{e-15}{\Msun}$ (purple), $\qty{e-9}{\Msun}$ (blue), and $\qty{e-3}{\Msun}$ (yellow). The frequency interval for the background is defined between the minimum and maximum values of $f_{\mathrm{peak}}$ in the contributing population. The sensitivity curves for \ac{lisa} (black dashed-dotted) and $\upmu$Ares (black dotted) are also illustrated.}
    \label{fig:asd_background}
\end{figure}
\begin{figure}[!t]
    \centering
    \includegraphics[width=\columnwidth]{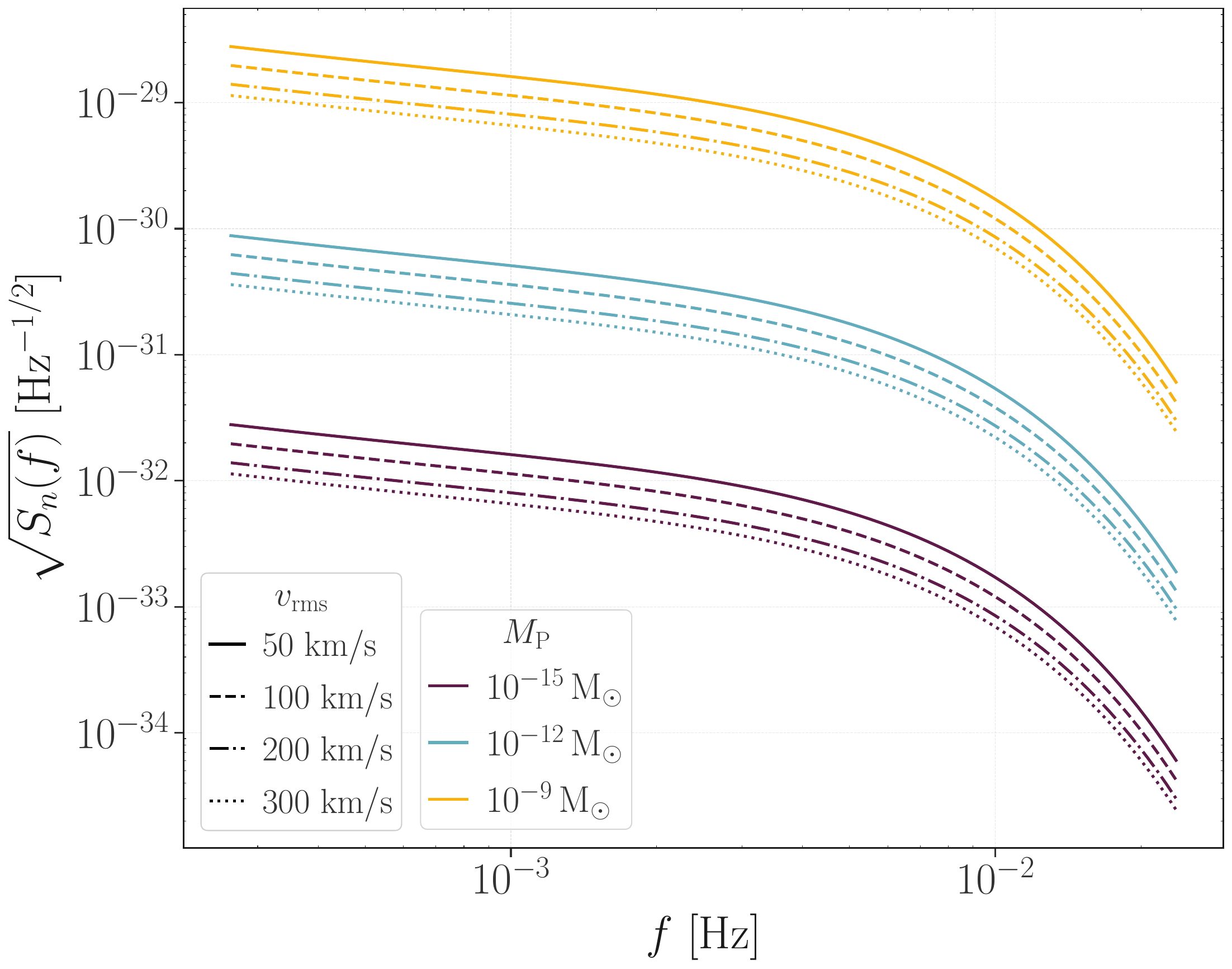}
    \caption{The frequency-dependent \ac{asd} for the popcorn gravitational wave background for the Milky Way (as illustrated in Fig.~\ref{fig:asd_background}) is show for three additional values of $v_{\mathrm{rms}}$. The variation in the \ac{asd} over these limiting values is less than an order of magnitude; illustrating the weak dependence of the signal on the choice of $v_{\mathrm{rms}}$.}
    \label{fig:vrms}
\end{figure}

In Fig.~\ref{fig:asd_background}, we present the \ac{asd} for a popcorn gravitational wave background generated by hyperbolic encounters of \acp{pbh} around \ac{sgr}, M31* and M87* compared to the sensitivity curves for \ac{lisa} and $\upmu$Ares \citep{robson_construction_2019}. The sensitivity curve for \ac{lisa} includes the confusion noise from unresolved galactic binaries estimated for a 4 year mission by \citet{cornish_galactic_2017}. The sensitivity curve for $\upmu$Ares is the detector noise only model, based on the proposal by \citet{sesana_unveiling_2021}. 

We plot curves for three representative \ac{pbh} masses of $\qty{e-15}{\Msun}$, $\qty{e-9}{\Msun}$, and $\qty{e-3}{\Msun}$ for each \ac{smbh}, which, as expected, scale in amplitude with the mass of the \ac{pbh}. Note that for the Milky Way, the event rate of \ac{pbh}-\ac{smbh} interactions is only above one per year for \ac{pbh} masses $M_{\mathrm{P}}\lesssim \qty{e-7}{\Msun}$. Above this, the signal is more appropriately treated as individual bursts rather than as a popcorn gravitational wave background, and this situation is discussed below.

As foreshadowed in Sec.~\ref{sec:geometry_event_rate}, the event rate, and thus the amplitude of the signal, depends relatively weakly on $v_\mathrm{rms}$, so we use a representative value of $\qty{100}{\kms}$ throughout. To verify this, Fig.~\ref{fig:vrms} shows the signal for the Milky Way from Fig.~\ref{fig:asd_background} for three \ac{pbh} masses ($\qty{e-9}{\Msun}$, $\qty{e-12}{\Msun}$, and $\qty{e-15}{\Msun}$) and $v_\mathrm{rms}$; $\qty{50}{\kms}$, $\qty{100}{\kms}$, $\qty{200}{\kms}$, and $\qty{300}{\kms}$, which span the typical range of central halo velocity dispersions. The \ac{asd} amplitude  varies less than an order of magnitude between the different $v_{\mathrm{rms}}$ values, justifying our choice in the context of a largely qualitative analysis. 

\begin{figure}[!t]
        \centering
        \vskip 2pt
    	\includegraphics[width=\columnwidth]{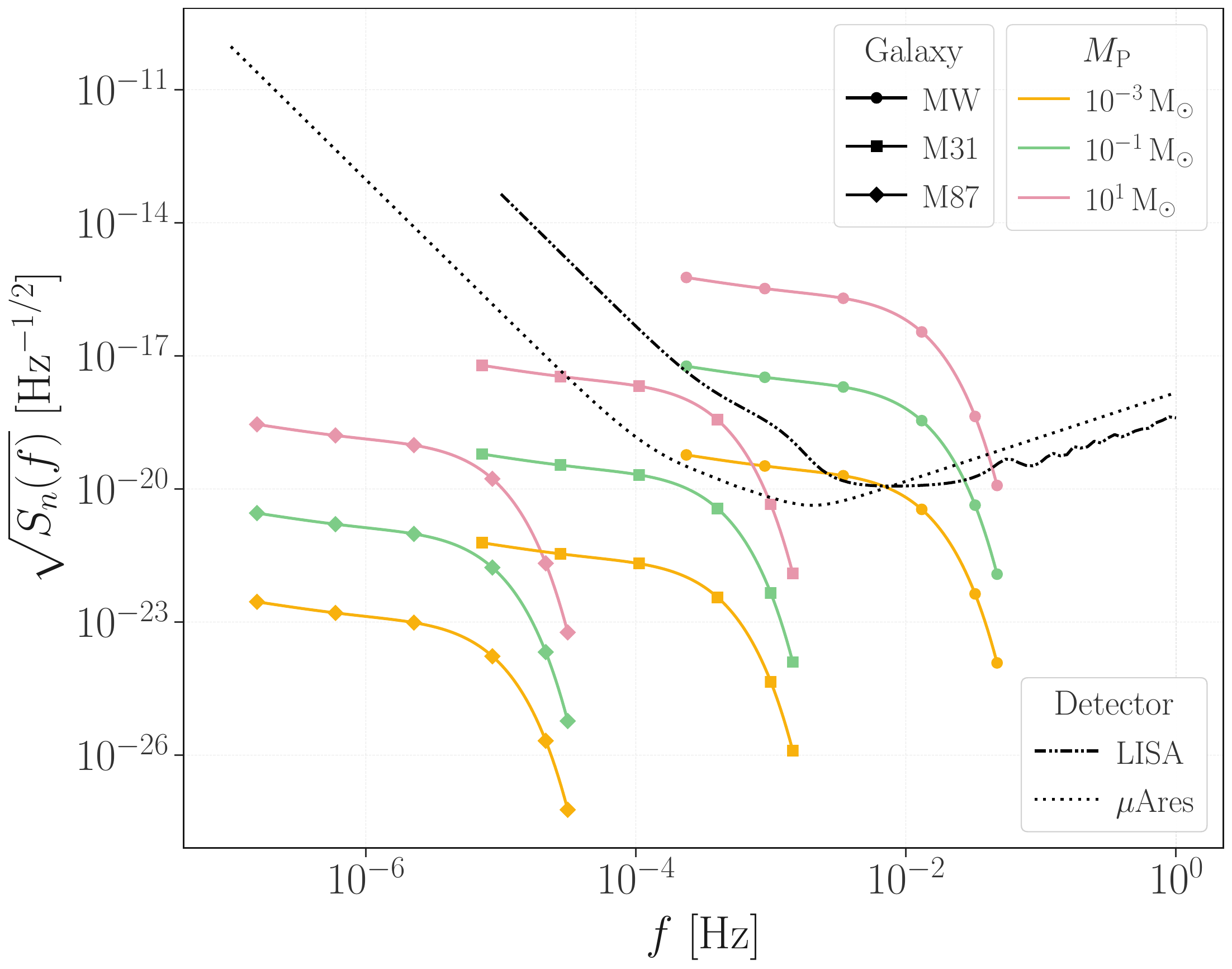}
    	\caption{The frequency-dependent \ac{asd} of a single hyperbolic encounter of a \ac{pbh} with \ac{sgr} (circles), M31* (squares), and M87* (diamonds) at a periastron distance of $\sim\qty{1}{\Rs} $. The \ac{asd} for three \ac{pbh} masses ($\qty{e-3}{\Msun}$, $\qty{e-1}{\Msun}$ and $\qty{e1}{\Msun}$) are represented over a logarithmic frequency interval centred around $f_{\mathrm{peak}}$. The sensitivity curves for \ac{lisa} (black dashed-dotted) and $\upmu$Ares (black dotted) are also illustrated.}
        \vskip 4pt
        \label{fig:asd_single}
\end{figure}

Fig.~\ref{fig:asd_single} shows the \ac{asd} of an individual hyperbolic encounter for large \ac{pbh} masses, $\qty{e-3}{\Msun}$, $\qty{e-1}{\Msun}$, and $\qty{e1}{\Msun}$. For single events, the signal is a short-duration, broadband burst, centred around $f_{\mathrm{peak}}$. We compute the \ac{asd} over a logarithmic frequency interval around this peak frequency to capture the full frequency content. We choose a periastron distance of $\sim 1 \,\mathrm{R_s}$, corresponding to the strongest possible burst. These signals are obviously independent of the density of the dark matter halo and the event rate. Moreover, the popcorn event rate will underestimate the burst signal when individual events are rare, consistent with our results here. 

We note that we present the signals and detector curves in Fig.~\ref{fig:asd_background} and Fig.~\ref{fig:asd_single} in terms of \ac{asd}. However, unlike characteristic strain, the relative height of the signal above a detector noise curve is not a direct measure of its \ac{snr}. For hyperbolic encounters, the signal is spread over a range of frequencies, and the \ac{snr} depends on the frequency-integrated signal relative to the detector noise. Hence, even signals which lie below the noise curve of a detector over much of their bandwidth can still yield appreciable \ac{snr}.
\newpage
\section{Detectability}
\label{sec:detect}
To assess the detectability of a signal with a given detector we consider the \ac{snr}, which compares the expected amplitude of the signal to the one-sided noise power spectral density of the detector $S_n(f)$. For \ac{lisa}, $S_n(f)$ is generated using tools provided by \citet{cornishEXtremeGravityInstituteLISA_Sensitivity2026}, and includes the full expression for the signal response function, which has been computed numerically \citep{larsonSensitivityCurvesSpaceborne1999}. The one-sided noise spectral density for $\upmu$Ares is based on the \ac{lisa} noise model rescaled to the mission parameters given by \citet{sesana_unveiling_2021}. 
\subsection{Bursts}
\begin{figure}[t!]
        \centering
        \vskip 2pt
    	\includegraphics[width=\columnwidth]{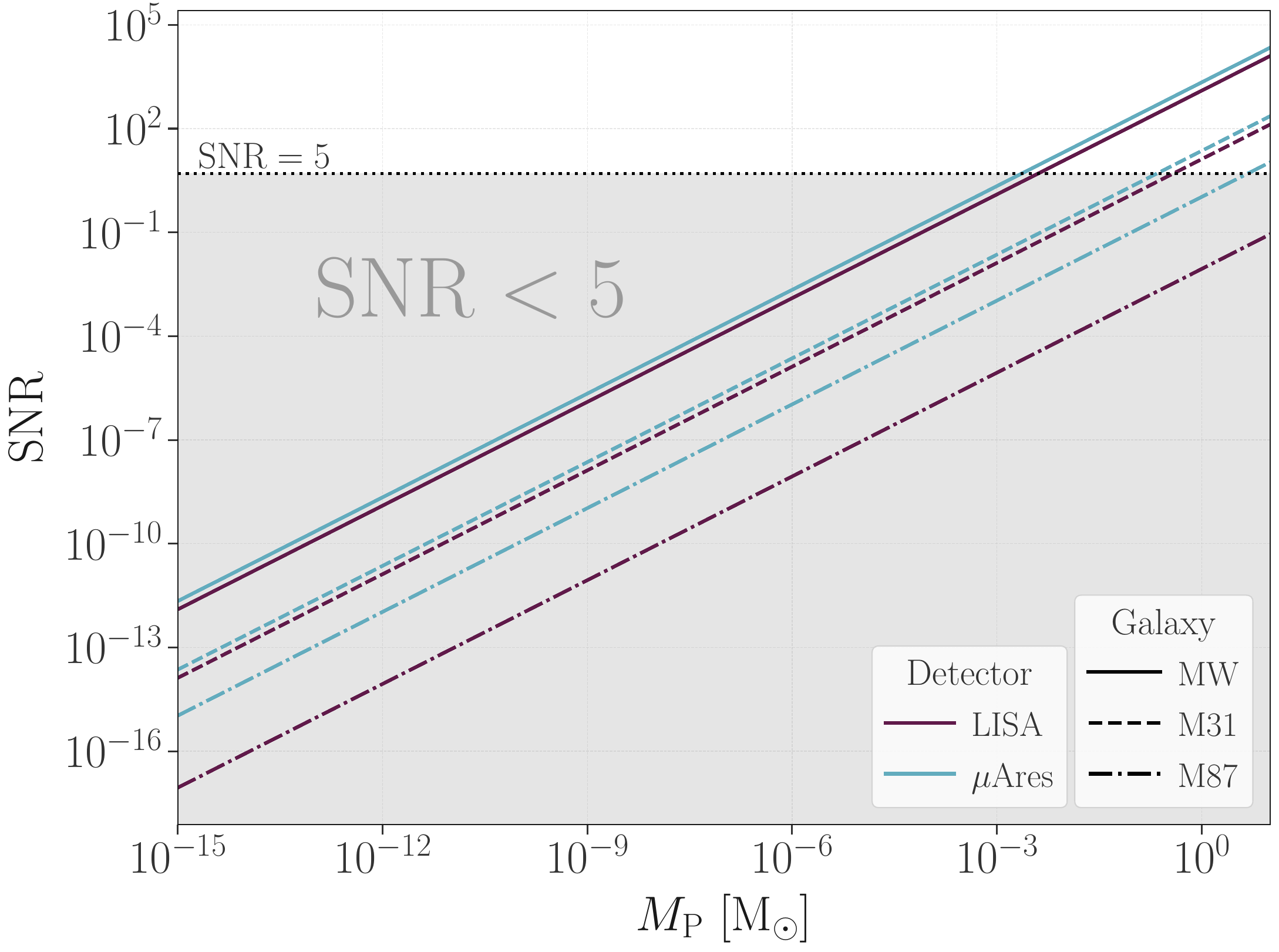}
    	\caption{The \ac{snr} in \ac{lisa} and $\upmu$Ares of a single hyperbolic encounter of a \ac{pbh} passing \ac{sgr} (solid), M31* (dashed), and M87* (dashed-dotted) for the full range of \ac{pbh} masses at a periastron distance of $\sim \qty{1}{\Rs}$. Labelled with a dotted line is $\mathrm{SNR} = 5$ and in grey is the region where $\mathrm{SNR} < 5$.}
        \label{fig:single_snr}
        \vskip 4pt
    \end{figure}
For a single gravitational wave burst, the \ac{snr} is given by \citet{smith_lisa_2019}
\begin{equation}
\mathrm{S N R}^{2} = 4 \int \frac{|\tilde{h}(f)|^{2}}{S_{n}(f)} \, \dd f \,.
\end{equation}
Fig.~\ref{fig:single_snr} shows the \ac{snr} as a function of $M_{\mathrm{P}}$ for hyperbolic encounters with \ac{sgr}, M31*, and M87* and periastron distance $\sim\qty{1}{\Rs}$ for \ac{lisa} and $\upmu$Ares. We indicate the region where $\mathrm{SNR}\geq5$, which we take as the threshold for detectability. For \ac{sgr}, \acp{pbh} with masses $M_{\mathrm{P}}\gtrsim \qty{e-2}{\Msun}$ produce signals with $\mathrm{SNR} > 5$ in both detectors. For M31*, only those with $M_{\mathrm{P}}\gtrsim \qty{e-1}{\Msun}$ reach the detection threshold for both detectors. In the case of M87*, $M_{\mathrm{P}} \gtrsim \qty{4}{\Msun}$ produces $\mathrm{SNR} > 5$ for $\upmu$Ares, but no bursts are visible to \ac{lisa}, due to their lower characteristic frequencies. 
\begin{figure}[!t]
        \centering
        \vskip 2pt
    	\includegraphics[width=\columnwidth]{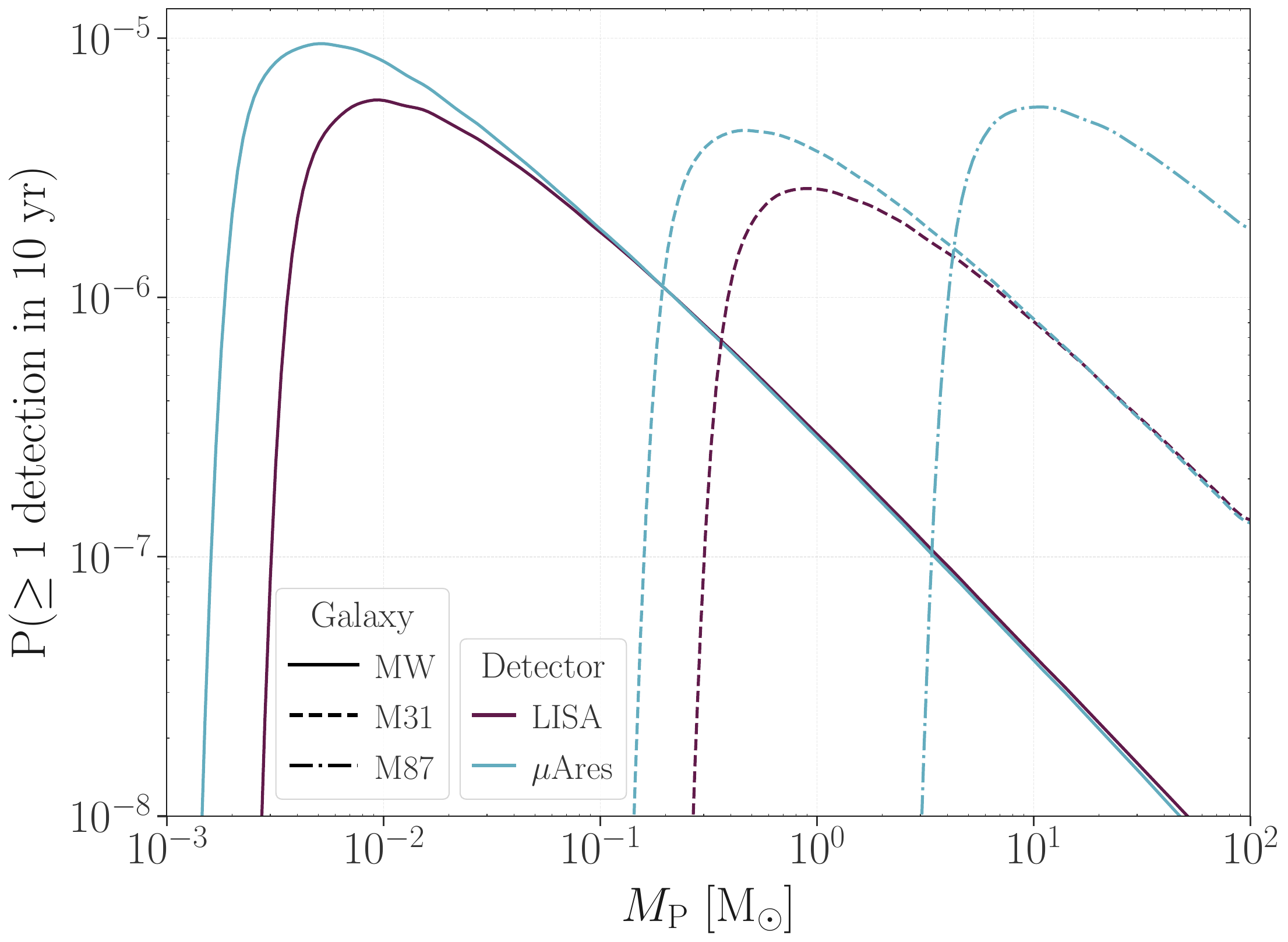}
    	\caption{Probability of detecting at least one gravitational wave burst signal from a hyperbolic \ac{pbh}-\ac{smbh} interaction over an observation period of ten years. The plotted curves are smoothed to reduce Monte Carlo sampling noise, while the detection probabilities are calculated directly from the simulated encounter populations. The most likely signals to be detected are from the Milky Way (solid) using $\upmu$Ares (blue), which is also able to detect interactions from the lowest \ac{pbh} mass. The probability for interactions from the Milky Way in \ac{lisa} (purple) is slightly lower, and the cutoff \ac{pbh} mass is slightly higher. The probability of a detection from M31 (dashed) and M87 (dashed-dotted) are comparable for $\upmu$Ares, however, for M87, only $\upmu$Ares is able to detect \ac{pbh}-\ac{smbh} interactions, with the cutoff mass being above $\qty{1}{\Msun}$.}
        \vskip 6pt
        \label{fig:probability}
\end{figure}

Unfortunately, if the \ac{pbh} mass is large enough for individual events to be detectable, the rates will be very low, even if the \acp{pbh} abundance is assumed to be unity. Treating the encounter rate as Poissonian, the probability of detecting at least one event, given our detection threshold of $\mathrm{SNR} \geq 5$, is $1 - \exp(-\lambda)$, where $\lambda$ is the expected number of detectable encounters over the operational lifetime of the detector.

Fig.~\ref{fig:probability} shows the probability of detectable events for the \ac{nfw} profile over 10 years of observations.\footnote{This is the nominal $\upmu$Ares mission and the potential extended \ac{lisa} mission, beyond its nominal 4 year lifetime.} %
For the largest \ac{pbh} masses, the \ac{snr} will exceed 5 for periastron distances $> 1 \, \mathrm{R_s}$, meaning that a greater number of trajectories can yield a detectable signal for larger black holes, but the overall flux scales inversely with the mass. For small $\lambda$, the probability of a detection scales linearly with the event rate, which is itself proportional to the abundance of \acp{pbh}. The sharp cutoff in the detection probability at lower masses comes from the chosen \ac{snr} threshold. In addition, any events near M87 are invisible to \ac{lisa}, given their relatively low frequency. The cutoff mass for M87 with $\upmu$Ares is $M_{\mathrm{P}} \sim \qty{4}{\Msun}$, so we have extended the \ac{pbh} mass range up to $\qty{e2}{\Msun}$ to capture the full curve. 
Given our assumed populations, it is clear the odds of  detectable events are very small. The most likely \ac{pbh} mass to be detected from the Milky Way is $M_{\mathrm{P}}\sim \qty{e-3}{\Msun}$, with a $\upmu$Ares detection slightly more likely than \ac{lisa}. 
\subsection{Popcorn Background}
For lower-mass \acp{pbh}, assuming an \ac{nfw} profile, the event rate can be as high as several encounters per second, resulting in an incoherent superposition of signals too weak to detect individually, or a so-called popcorn background. In this case, the \ac{snr} grows with the total observation time $T_{\mathrm{obs}}$ \citep{allen_detecting_1999, thrane_sensitivity_2013}
\begin{equation}
\mathrm{S N R}^{2} = 4 T_{\mathrm{obs}} \int \frac{|\tilde{h}(f)|^{2}}{S_{n}(f)} \, \dd f \,.
\end{equation}

\begin{figure}[!t]
        \centering
    	\includegraphics[width=\columnwidth]{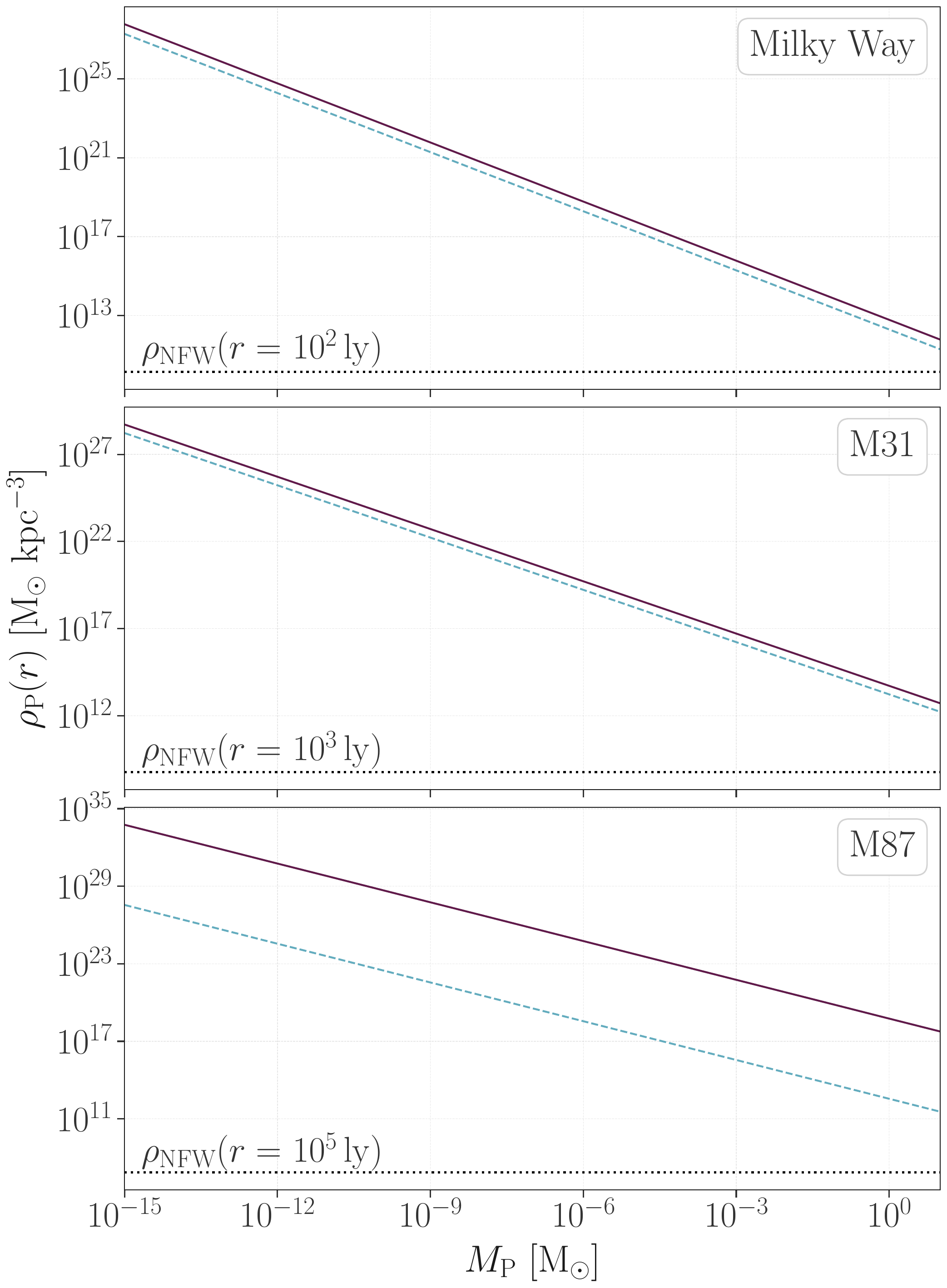}
    	\caption{The density of \acp{pbh} $\rho_{\mathrm{P}}$ required as a function of \ac{pbh} mass for a popcorn gravitational wave background from hyperbolic encounters to have an \ac{snr} of 5 in \ac{lisa} (solid purple) and $\upmu$Ares (dashed blue), assuming a ten year observation period. The assumed density of the \ac{nfw} profile at $r_i$ for each galaxy is indicated by a dotted line.}
        \label{fig:pbh_density}
\end{figure}

The detectability of a gravitational wave background from hyperbolic flybys of lighter \acp{pbh} depends strongly on the halo density profile. We assess the density of \acp{pbh} needed for a detectable gravitational wave background, again assuming a threshold $\mathrm{SNR} \geq 5$. Fig.~\ref{fig:pbh_density} shows the density of \acp{pbh} $\rho_{\mathrm{P}}$ needed at $r = \qty{e2}{\ly}$ for the Milky Way, $r = \qty{e3}{\ly}$ for M31, and $r = \qty{e5}{\ly}$ for M87, that would generate detectable signals in \ac{lisa} and $\upmu$Ares. The estimated density of the \ac{nfw} profile at the relevant distance for each galaxy is labelled (dotted line) and this is far below the threshold in all cases.  

\section{Conclusions}
\label{sec:conclusion}

This analysis yields a very pessimistic assessment of the ability of gravitational wave signals generated by their interactions with \acp{smbh} to put useful bounds on a possible \ac{pbh} population that contributes to galactic dark matter. However, the analysis is still informative. Firstly, it highlights between the single-burst and popcorn signals expected in different mass ranges and it draws attention to the possibility of detections from M31* and M87*; the former in particular has been largely overlooked in previous treatments of \ac{pbh}-\ac{smbh} interactions. In addition, while the signal we find is well below the threshold of detectability, \ac{pbh} dark matter will necessarily have hyperbolic encounters with central \acp{smbh}, so estimating its visibility is a necessary task. 

It is also very possible that we have overestimated the signal -- for example, by using an \ac{nfw} rather than a Burkert profile. On the other hand, there are also several ways in which our estimate might be an underestimate. For example, we have assumed the dark matter is smoothly distributed, but it has been suggested that \acp{pbh} could form bound clusters \citep{meszarosPrimevalBlackHoles1975}. In this case, the local number density at galactic centres could be enhanced considerably from the \ac{nfw} profile, and the constraints on the abundance of \acp{pbh} from microlensing observations could be significantly relaxed \citep{belotskyClustersPrimordialBlack2019}. 

We note that we assumed a monochromatic \ac{pbh} mass distribution. This is likely to be unrealistic given that most \ac{pbh} formation mechanisms naturally yield an extended mass function \citep{carrPrimordialBlackHole2017}. Depending on the shape of this distribution, a contribution from a range of \ac{pbh} masses could either enhance or suppress the overall signal.

This analysis offers generic insight into the detection of \ac{pbh}-\ac{smbh} signals -- with currently proposed detectors, the Milky Way offers the best probe of sub-solar mass \acp{pbh}. Moreover, proposals such as AMIGO \citep{Baibhav:2019rsa} and the Decihertz Observatory \citep{Sedda:2021yhn} would significantly improve on the sensitivity of \ac{lisa} in these frequency ranges. In addition, while \ac{lisa} does not have excellent sky-localisation for transient sources, a pair of well-separated space-based detectors would provide a useful baseline. Beyond a $\sqrt{N}$ improvement in \ac{snr}, the fact that these signals come from known locations on the sky would allow a network of detectors to offer further leverage. Interestingly, M31 interactions are accessible to both \ac{lisa} and $\upmu$Ares, but do not seem to have received significant consideration in previous work. Finally, this work highlights that point sources moving near M87* will emit gravitational waves at frequencies lower than any planned space-based interferometer.

Lastly, we should also stress that our focus on unbound orbits sets a lower bound on any possible signal, and our estimates of the dark matter density (particularly for M87) are taken from points some distance from the actual centre of the galaxy. Moreover, we are ignoring the contribution of a bound \ac{pbh} population, which can significantly enhance the dark matter density near an \ac{smbh} \citep{gondolo_dark_1999}. Fully investigating these possibilities represent clear directions for future work.

\section*{Acknowledgments}

The authors gratefully acknowledge support
from the Marsden Fund Council grant MFP-UOA2131, funded
by the New Zealand Government and managed by the Royal
Society Te Apārangi. This work was supported by the French CNRS International Research Project (IRP) OG-Science FR-NZ. NC has received financial support from the French Agence Nationale de la Recherche.

\bibliographystyle{aasjournal}
\bibliography{oja_template}

\begin{appendix}

\section{Appendix 1: Derivation of periastron}
\label{ap:r_min}
We review the derivation of the distance of closest approach, or periastron distance $r_{\mathrm{min}}$, for a \ac{pbh} passing an \ac{smbh} on an unbound, hyperbolic orbit.  From Fig.~\ref{fig:hyperbolicgeometry} we see that the impact parameter $b$ and angle $\alpha$ satisfy
\begin{equation}
    b = r\sin\alpha\,,
    \label{eq:impact_param_alpha}
\end{equation}
and
\begin{equation}
    \tan\alpha = \frac{b}{\sqrt{r^2 - b^2}}\,.
\end{equation}
In an inverse–square potential, the Cartesian equation of the orbit can be written as \citep{kibble2004classical}
\begin{equation}
    \frac{(x-ae)^2}{a^2} - \frac{y^2}{b^2} = 1\,,
    \label{eq:cartesian}
\end{equation}
where $e$ is the eccentricity, and
\begin{equation}
    a = \frac{l}{e^2 - 1} = \frac{|k|}{2E} = \frac{GM_{\mathrm{S}}}{v_0^2}
\end{equation}
is the semi-major axis. The impact parameter is 
\begin{equation}
    b^2 = al = \frac{J^2}{2M_{\mathrm{P}}E}\,,
\end{equation}
where $E = \frac{1}{2}M_{\mathrm{P}}v_0^2$ is the total energy of the \ac{pbh} at infinity, where $v_0$ is the asymptotic velocity. The coupling constant is $|k| = GM_{\mathrm{S}}M_{\mathrm{P}}$  and $l = J^2 / GM_{\mathrm{S}}M_{\mathrm{P}}^2$ is the length parameter. The angular momentum of the \ac{pbh} is $J = M_{\mathrm{P}}v_0b$. The distance of closest approach occurs when $y =0$, reducing Eq.~(\ref{eq:cartesian}) to $(x-ae)^2 = a^2$. Placing the \ac{smbh} at the origin and following the convention $a > 0$, yields
\begin{equation}
    x_{\mathrm{min}} -ae = -a\,,
\end{equation}
so that 
\begin{equation}
    r_{\mathrm{min}} = a(e-1) = \frac{GM_{\mathrm{S}}}{v_0^2}(e-1)\,.
    \label{eq:r_min}
\end{equation}
We can express the eccentricity as a function of the impact parameter and the initial velocity
\begin{equation}
    e^2 = \frac{v_0^4 b^2}{(GM_{\mathrm{S}})^2} + 1\,.
    \label{eqn:eccsquared}
\end{equation}
Substituting the above into Eq.~(\ref{eq:r_min}) gives 
\begin{equation}
    r_{\mathrm{min}} = \sqrt{b^2 + \left(\frac{GM_{\mathrm{S}}}{v_0^2}\right)^2} - \frac{GM_{\mathrm{S}}}{v_0^2}\,.
\end{equation}

\end{appendix}

\end{document}

%% file: glossary.tex
\DeclareAcronym{nfw}{
  short = NFW ,
  long  = Navarro--Frenk--White
}

\DeclareAcronym{mbh}{
  short = MBH ,
  long  = massive black hole
}

\DeclareAcronym{smbh}{
  short = SMBH ,
  long  = supermassive black hole
}

\DeclareAcronym{sgr}{
  short = Sgr~A* ,
  long  = Sagittarius~A*
}

\DeclareAcronym{pbh}{
  short = PBH ,
  long  = primordial black hole
}

\DeclareAcronym{lisa}{
  short = LISA ,
  long  = Laser Interferometer Space Antenna
}

\DeclareAcronym{snr}{
  short = SNR ,
  long  = signal-to-noise ratio
}

\DeclareAcronym{wimp}{
  short = WIMP ,
  long  = weakly interacting massive particle
}

\DeclareAcronym{lvk}{
  short = LVK ,
  long  = LIGO-Virgo-KAGRA
}

\DeclareAcronym{m87}{
  short = M87 ,
  long  = Messier 87
}

\DeclareAcronym{cdm}{
  short = CDM ,
  long  = cold dark matter
}

\DeclareAcronym{psd}{
  short = PSD ,
  long  = power spectral density
}

\DeclareAcronym{asd}{
  short = ASD ,
  long  = amplitude spectral density
}